\documentclass[fleqn,10pt]{wlscirep}
\usepackage[utf8]{inputenc}
\usepackage[T1]{fontenc}
\usepackage{enumerate}
\usepackage[skip=0.5pt]{subcaption}
\usepackage[width=.9\textwidth]{caption}
\usepackage{graphicx}
\usepackage{multirow}
\usepackage{amssymb}
\usepackage{amsmath}
\usepackage{bbm}

\title{Inferring urban polycentricity from the variability in human mobility patterns}

\author[1,2 *]{Carmen Cabrera-Arnau}
\author[2]{Chen Zhong}
\author[2]{Michael Batty}
\author[3]{Ricardo Silva}
\author[4]{Soong Moon Kang}

\affil[1]{Department of Geography and Planning, University of Liverpool, Liverpool,  United Kingdom}
\affil[2]{Centre for Advanced Spatial Analysis (CASA), University College London, London, United Kingdom}
\affil[3]{Department of Statistical Science, University College London, London, United Kingdom}
\affil[4]{School of Management, University College London, London, WC1E 6BT, United Kingdom}

\affil[*]{Corresponding author: c.cabrera-arnau@liverpool.ac.uk}

\begin{abstract}
The polycentric city model has gained popularity in spatial planning policy, since it is believed to overcome some of the problems often present in monocentric metropolises, ranging from congestion to difficult accessibility to jobs and services. However, the concept `polycentric city' has a fuzzy definition and as a result, the extent to which a city is polycentric cannot be easily determined. Here, we leverage the fine spatio-temporal resolution of smart travel card data to infer urban polycentricity by examining how a city departs from a well-defined monocentric model. In particular, we analyse the human movements that arise as a result of sophisticated forms of urban structure by introducing a novel probabilistic approach which captures the complexity of these human movements. We focus on London (UK) and Seoul (South Korea) as our two case studies, and we specifically find evidence that London displays a higher degree of monocentricity than Seoul, suggesting that Seoul is likely to be more polycentric than London.
\end{abstract}

\begin{document}

\flushbottom
\maketitle

\thispagestyle{empty}

\vspace{-4mm}

\section{Introduction}

\subsection{Measuring monocentricity}

People in cities interact with their environment by developing urban land for different socioeconomic activities. The way in which land use is located and arranged within a city is either a result of self-organising mechanisms over the course of time or as a result of specific interventions through different varieties of urban planning, at different spatial scales. In this context, the configuration is usually referred to as urban structure and through the study of such structures, we can learn more about the spatial behaviours of the societies that, over time, have built them. Moreover, urban structure also plays an important role in shaping the present and future, given its impact on different socioeconomic features such as mobility, access to jobs, social mixing, heterogeneity, segregation, deprivation, urban efficiency, and sustainability.

The simplest form of urban structure corresponds to the monocentric city, where socioeconomic activity is localised in a unique central region. In practice, monocentric cities facilitate the accumulation of social interactions and innovation, and consequently give rise to economies of agglomeration characterised by increasing returns to scale \cite{Anas88, Rosenthal01}. However, monocentric cities are also subject to heavy tidal flows on the transport facilities during peak hours, severe congestion and disproportionally high rents close to the centre \cite{Ahfeldt13, Anas2000, YueHuai21}. The monocentric structure of cities prevailed until the industrial revolutions led to new forms of transport that broke the bounds of compact cities. Consequently, monocentric cities have gradually decentralized, transforming into complex hierarchies of different kinds of centres, neighbourhoods and sprawling structures that are tied together by a multiplicity of transport and information systems \cite{Zhong14}. Yet, explanatory models of urban structure based on a monocentric approach are still used due to their simplicity and formal analytical elegance. Their validity, however, should be questioned both in terms of the theoretical assumptions needed for the formulation of the models \cite{Wheaton79, Griffith81, Berry93} and from the point of view of public policy \cite{ArribasBel14}, since most plans for future cities have long abandoned the idea of the monocentre.

Polycentricity has therefore become the focus of much spatial policy \cite{Green07}, since it is believed that urban dwellers in polycentric cities might benefit from congestion relief in comparison with their monocentric counterparts \cite{Brinkman16} and from increased accessibility to jobs and services, which may translate in higher rent and housing prices all across the city, but also in more time-efficient and cost-efficient travel. Despite the raise in popularity of the idea of polycentric development, it remains a rather fuzzy concept as it seems to mean different things to different actors and on different scales \cite{Davoudi03, Green07, Kloosterman01, Meijers08, Rauhut17}. The lack of a concise and coherent definition raises an issue: how to measure polycentricity? If we do not know what to measure, we simply cannot measure it \cite{Green07, Meijers08, Rauhut17}. In this work, instead of attempting to answer the ill-defined question `to what extent is a city polycentric?', we provide an approach to analysing departures from a well-defined concept of monocentricity.

Despite the fuzziness in the definition of `polycentric city', there is a long tradition of theoretical research and empirical evidence surrounding the debate on monocentricity versus polycentricity. We will simply indicate recent work here such as that based on an analysis of data from US metropolitan areas by Arribas-Bel and Sanz-Garcia \cite{ArribasBel14}, which shows that monocentricity still retains a substantial influence on the intraurban structure of many metropolitan areas. This is despite the general consensus in the literature that modern cities above a certain size threshold become polycentric and that monocentricity is an older concept more appropriate to the city in history prior to the industrial revolution. In this sense, the concept might perhaps be somewhat obsolete when dealing with the real world. Additionally, the authors cited in \cite{ArribasBel14} find that there is no clear evolutionary trend in US cities towards polycentricity between 1990 and 2010. 

By contrast, Alidadi and Dadashpoor \cite{Alidadi18} analyse data from Iran to find that a monocentric model is not able to explain the spatial distribution of employment in Tehran while the main core has been losing its importance with the passage of time. Li \cite{Li20} draws upon fine-grained LandScan population data corresponding to 286 Chinese cities to find that in general, urban spatial structure has become more polycentric as well as more concentrated (i.e. with a higher share of their population living in the centres) while these changes have usually resulted in population and economic productivity growth.

There are other studies that find evidence for mixed types of urban structure. Hajrasouliha and Hamidi \cite{Hajrasouliha17} base their study on three typologies of urban structure: monocentricity, polycentricity, and generalized dispersion. When analysing the spatial structure of employment data from 356 US metropolitan regions, they find that mixed typologies of urban structure outnumber the three ``pure'' ones by almost four to one. They also find that polycentricity is somewhat more common than monocentricity. Similarly, in \cite{Sweet17}, Sweet et al use cross-sectional data to estimate the relative strengths of monocentricity, polycentricity, and dispersion for characterizing Canadian cities. Their results indicate that elements of each model are evident, but each tends to dominate in different contexts. When focusing on Montreal, Toronto, and Vancouver, their results imply that accessibility, municipal competition, and globalization play a role in shaping urban structure.

\subsection{Using new forms of data}

In the past, most empirical studies of urban structure based their conclusions on data associated with the spatial distribution of employment or population, obtained largely from traditional sources of direct observation by questionnaires such as surveys, censuses and administrative records. The rationale behind this choice of datasets is that they are comprehensive and representative and have the potential to uncover where city dwellers conduct most of their socioeconomic activity. It has only been in recent years that the focus has been turned to alternative data sources, which can offer real-time and easy-access records at very small spatial scales. In particular, the locations that people choose to visit at different times of the day or week are very much conditioned by the spatial structure of the city and at the same time, the complexity of human movements shapes the usage of urban space and the arrangement of resources \cite{Zhong14}. Therefore, the study of patterns in human mobility through alternative data sources can help us understand the travel behaviour of city dwellers and it can also help us uncover the socio-economic features of urban structure. 

For example, recent studies have used data derived from social media platforms as well as location tracker devices in mobile phones in order to understand the spatial structure of cities based on the places visited by the users \cite{Tu17, Yin17, Yang18, Zhu20, Xiao21, Liu21, Ponce-Lopez21, Miao21, Zhenglin22}. Taxi trajectory data is another alternative source of data that has gained in popularity in recent years as a means to uncover information about urban structure \cite{Zhou15, Nie21, Li21, Choi22}. Taxi trajectory data not only has the potential to reveal the characteristics of human movement within the city, but also real-time traffic status as well as potential social inequalities. A third alternative source of data is that derived from smart travel cards or simply, smart cards. Like the other sources of alternative data already mentioned, smart card data offers information regarding daily human activities at high resolution, both in the spatial and temporal domains and consequently, it has been used to explore urban structure \cite{Roth11, Maeda19, Yang19, Tang20, Nilufer21, Zhang21}. Here, we will focus on the latter type of new data sources that record such movements. In our case, this is smart card data from the automatic fare collection system in London's and Seoul's public transport, which contains information about the origin, destination and time at which each individual journey occurs. 

\subsection{Aim and contribution}\label{sec:aim}

Our aim is to provide a novel approach to model the extent to which a city departs from the monocentric structure by considering the variability inherent in human mobility patterns, and by avoiding the fuzziness in the concept of polycentricity. To investigate the applications of this approach, we consider two case studies using high spatiotemporal resolution data derived from smart travel cards corresponding to London, United Kingdom, and Seoul, South Korea.

Our methodology first considers the frequency distribution of the length of journeys terminating at each station in the public transport system of a given city on a typical weekday. We define the ``nucleus'' of each city as the station representing a hypothetical centre. We also consider the network structure of the public transport system in order to measure the length of the journeys by the network distance between stations. We then introduce Poisson mixture models as a statistical approach to describe the frequency distribution of the length of the journeys terminating at each station in the transport system. The Poisson mixture models enable us to capture the variability in the human mobility patterns reflected in urban structure, which in real cities includes a blend of features which characterise both monocentric and polycentric cities.

Next, we state what we call the \textit{monocentric hypothesis}: ``If a city was perfectly monocentric, the expected length of the journeys taken to a given station, except for the nucleus itself, would be equal to the length of the shortest path between the nucleus and the destination station''. In this hypothetical scenario, the nucleus would be the only centre for socioeconomic activity in the city, and consequently, a typical journey terminating at a given station other than the nucleus would have its origin at the nucleus. Journeys whose destination is the nucleus would have their origin at various locations across the public transport system.  

In reality, cities and urban mobility patterns are more complex than stated by the \textit{monocentric hypothesis}, so quantifying deviations from this idealistic behaviour enables us to understand the extent to which a city departs from monocentricity, or in other words, it enables us to indirectly infer its degree of polycentricity.

Therefore, the main contribution of this analysis is a solution to the problem of quantitatively describing the degree of monocentricity of a city. Our data-driven approach based on mixture models considers the complexity of urban space since these models are able to capture the variability in human movements that arises as a result of sophisticated forms of urban structure. Instead of considering discrete typologies of urban structure (e.g. monocentric and polycentric), the method proposed here conceptualises urban structure typologies as a spectrum, where monocentricity is an idealistic extreme. According to the observed patterns of human mobility, we specifically find evidence that London displays a higher degree of monocentricity than Seoul, suggesting that Seoul is likely to be more polycentric than London.

The rest of the paper is organized as follows. In Section 2, we describe in detail the data sets used for the analysis and how the data has been processed. Section 3 is dedicated to the methodology which we followed for the analysis. We explain how we conceive of the public transport system as a complex network. We also introduce the probabilistic modelling framework and mixture models. In Section 4, we present the results of the analysis corresponding to the two case studies of London and Seoul. We provide some concluding remarks and points of discussion in Section 5. We also included Supplementary Information with some additional results to support our findings and conclusions.

\section{Methodology}

\subsection{Data and notation}

The Oyster card in London and the T-money card in Seoul are automatic fare collection systems that record the place and time when a traveller enters and exists the public transport system by tapping in and out with their card. In 2012, more than 80\% of all journeys on public transport in London were made using Oyster card whereas in Seoul 98.9\% of all journeys on public transport were made using T-money in 2013. Our study is based on tap-in and tap-out records of Oyster and T-money card but we only use tap-outs because we consider that these are a better indication of employment locations in the morning peak whereas tap-outs in the evening peak are more likely to involve trips that involve travel to entertainment. The taps that we used are thus aggregated at a temporal resolution of 1 hour. For London, we used data recorded 24 hours on 5 weekdays between January 20 and 24 in the year 2014, and for Seoul, 4 weekdays between December 17 and 21 in the year 2012. We excluded the data from Wednesday, December 19, because it was a presidential election day in Korea and regular travel patterns were disrupted. 

We process the data to obtain the number of journeys of a given length which terminate at each station on a typical weekday. Data for a typical weekday is obtained by averaging the number of counts across all the weekdays included in the raw data set and rounding the average value to the closest integer. The total daily average count of journeys was 3.22 million encompassing 382 stations in London and 5.96 million for 512 stations in Seoul.

For the subsequent parts of the analysis, we introduce the following notation. Each of the $N$ stations in each city is symbolised by $S_i$, with $i=1,...,N$. Station $S_i$ is the destination of $M_i$ journeys so it has $M_i$ tap-outs. The length of the $l$th journey terminating at station $S_i$ is symbolised by $L_i^l$, where $l$ is an index over the $M_i$ journeys to $S_i$ and therefore, $l=1,...,M_i$. 

\subsection{The transport system as a complex network}

We define an undirected simple network $G = (V,E)$ as an abstract conceptualisation of the public transport system in London or Seoul. The network $G$ is formed by the set of $N$ vertices or nodes $V$ and the set of edges $E$. The $i$th node of the network $G$ corresponds to station $S_i$ in the public transport system. An edge is present between two nodes $i$ and $j$ if there is at least a line of transport that provides a direct connection between the stations $S_i$ and $S_j$. The distance between stations $S_i$ and $S_j$ is symbolised by $d_{ij}$ and is defined as the sum of the distances associated with the minimum number of edges that need to be traversed in order to travel from $S_i$ to $S_j$. The length of a journey between two stations is defined as the  distances associated with the number of edges that are traversed from the origin to the destination nodes, but here we assume that the length of a journey between stations $S_i$ and $S_j$ is equal to the distance $d_{ij}$, i.e. we assume that, from all possible trajectories from $S_i$ to $S_j$, passengers always choose the one involving the fewest stops. 

For the next steps of the analysis, we establish a hypothetical centre in the network, which we call the ``nucleus''. Different notions of centrality can be considered, although not all of them are suitable for our analysis. For example, if we considered that the nucleus is the closest station to the geographical centroid of the city, then the definition of centre would depend on the physical boundaries of the city region, which in turn, can be established according to a variety of different criteria. If instead, we considered a measure of centrality based on the topology of the network, such as the betweenness centrality of each node, then the nucleus in Seoul would be Wangsimni station, which does not necessarily represent what many Seoulites would consider to be a central region of Seoul. Similarly, a measure of centrality based on traffic flows may also not coincide with what most people consider to be the centre. For these reasons, we opt for a somewhat arbitrary choice of nucleus based on what is popularly considered to be a central area: Piccadilly Circus station in London and City Hall station in Seoul. We assign the index $i=1$ to the station corresponding to the nucleus in each city. To counter the arbitrariness of these choices of nucleus, we provide a sensitivity test of the results of our analysis. This can be found in the Supplementary Information, in the section titled `Sensitivity analysis for different choices of nucleus'.

\begin{figure}[ht]
\centering
    \includegraphics[width=0.67\textwidth]{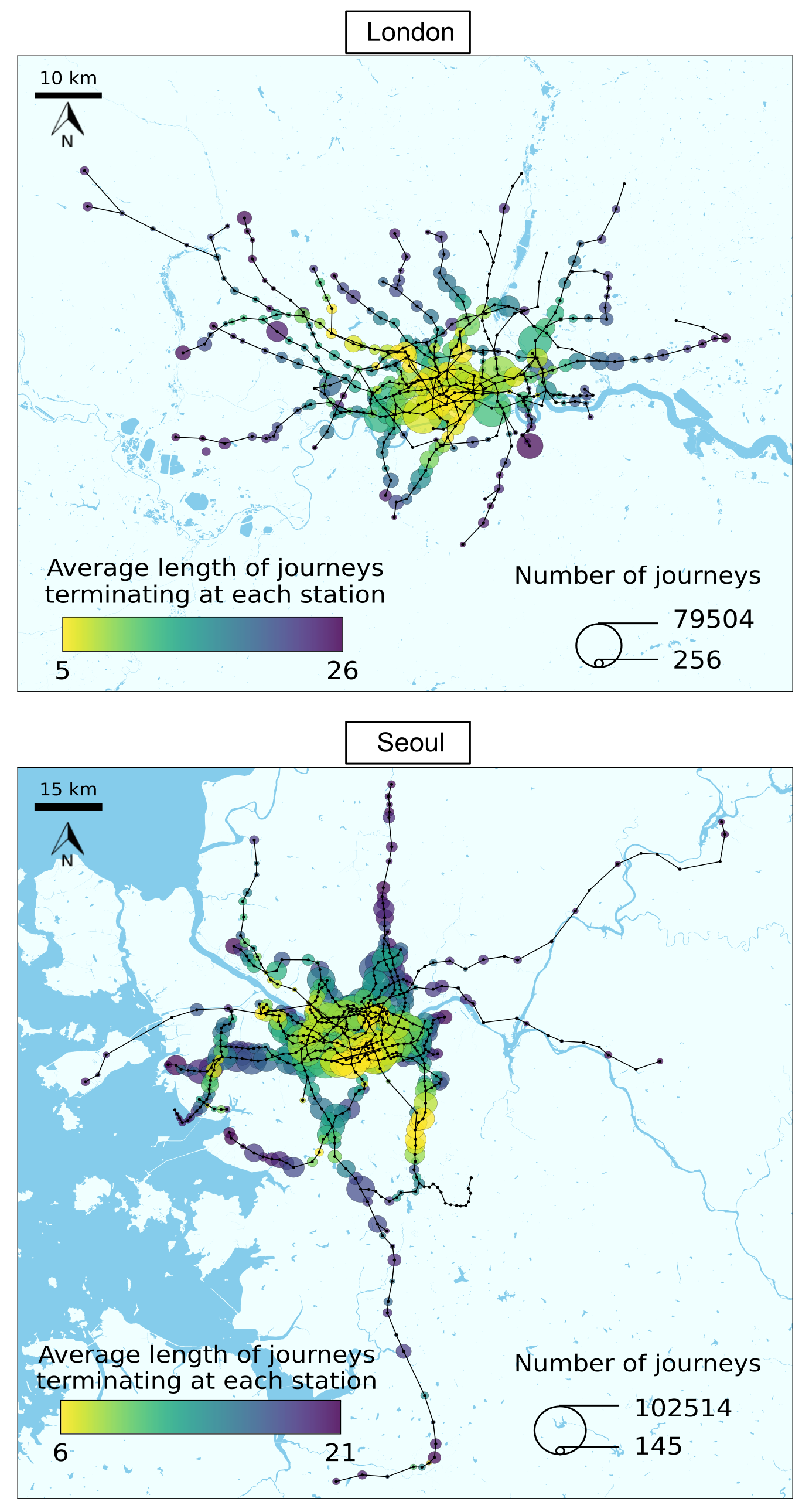}
    \vspace*{4mm}
    \caption{Map of London's and Seoul's public transport systems.}\label{fig:map}
\end{figure}

Figure \ref{fig:map} shows the physical layout of London and Seoul's public transport system, with the lines and stations that are included in the data sets. The transport lines are traced simply as straight lines to show the topology of the network. The colour of the nodes corresponds to the average length of the journeys terminating at $S_i$, symbolised by $\bar{L}_i$ and the size represents the number of journeys $M_i$ reaching each station $S_i$.

\subsection{Modelling the distribution of journey lengths}

In this section, we introduce a probabilistic approach to model the frequency distribution of journey lengths on a typical weekday. We regard $L_i$ as a discrete random variable denoting the length of journeys whose destination is station $S_i$. For each station, our data set gives $M_i$ realisations of $L_i$, so the observed length of the $l$th journey, symbolised by $L_i^l$, would correspond to the $l$th realisation of $L_i$. The true probability distribution of random variable $L_i$ is unknown, however, its empirical probability density function, denoted by $\hat{f}_i$, can be obtained from the observed data as 

\begin{equation}\label{empirical_pdf}
    \hat{f}_i(L_i = h) = \frac{1}{M_i}\sum_{l=1}^{M_i}\mathbbm{1}_{L_i^l=h},
\end{equation}

with $h\in \mathbb{N}$. In equation \eqref{empirical_pdf}, $\mathbbm{1}_{L_i^l=h}$ is an indicator function that takes the value 1 when $L_i^l = h$ and zero otherwise. Hence, the probability that the length of a journey with destination at station $S_i$ is equal to $h$ is approximated by $\hat{f}_i(L_i = h)$, computed as the number of observed counts of journeys of length $h$ terminating at $S_i$, divided by the total number of journeys terminating at $S_i$, i.e. $M_i$.

Under the \textit{monocentric hypothesis} stated in section \ref{sec:aim}, if a city was perfectly monocentric, the expected length of the journeys taken to a given station, except for the nucleus itself, would be equal to the length of the shortest path between the nucleus and the destination station. Hence, this null hypothesis can be expressed mathematically as
\eqref{hypothesis}
\begin{equation}\label{hypothesis}
    E[L_i] = d_{1i}
\end{equation}
for $i=2,...,N$, where $E[L_i]$ is the expected value of random variable $L_i$, which can be approximated by the sample mean $\hat{\mu}_i = \frac{1}{M_i}\sum_{l=1}^{M_i}\mathbbm{1}_{L_i^l}$. Taking this into account, the \textit{monocentric hypothesis} can be expressed as $\hat{\mu}_i = d{1i}$.

\begin{figure}[ht]
\centering
\leavevmode
    \hbox{\includegraphics[width=0.67\textwidth]{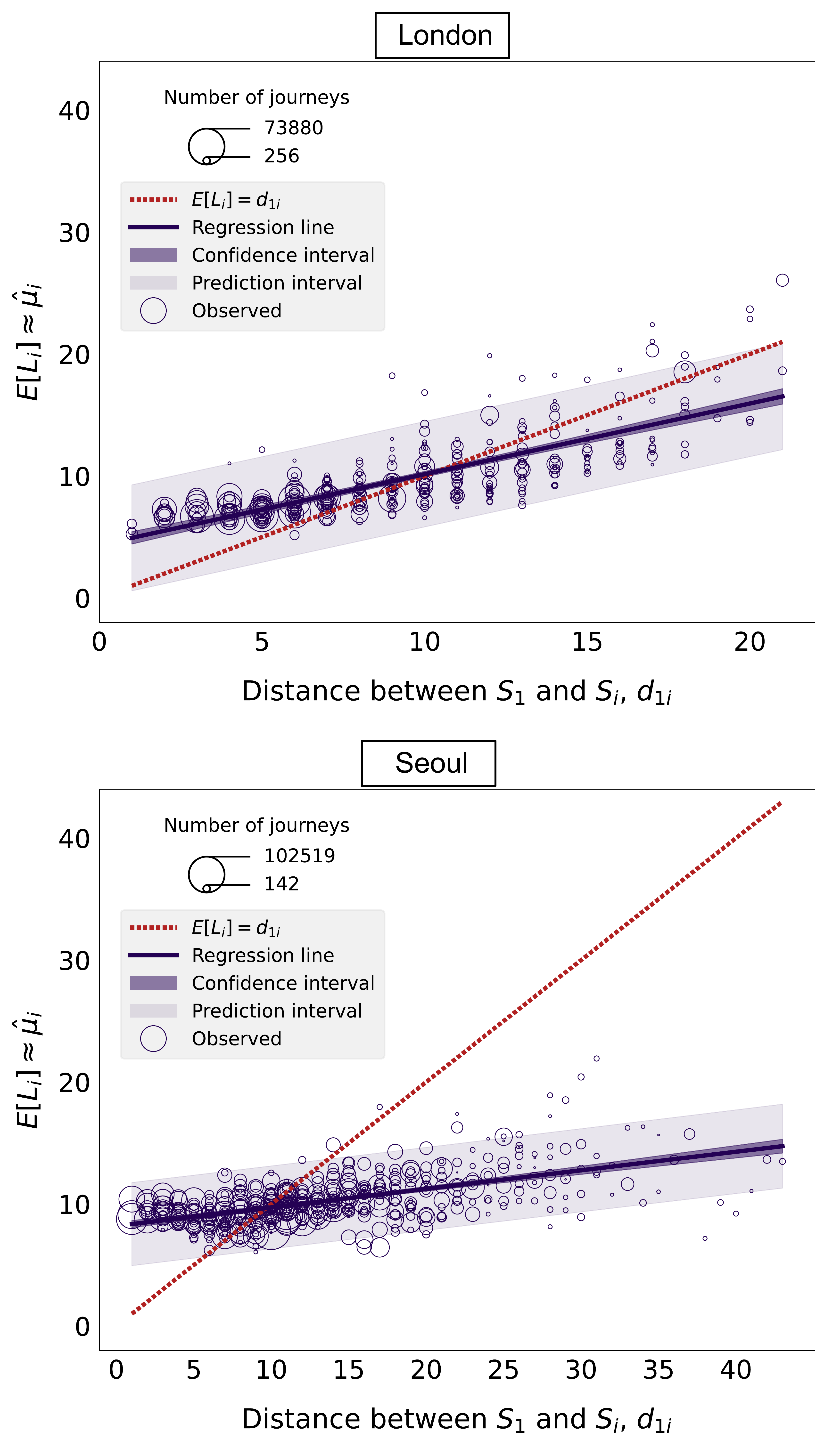}}
    \vspace*{3mm}
    \caption{Relationship between the mean of the distribution of journeys terminating at each station, $\hat{\mu}^p_i$ and $\hat{\mu}^d_i$, and the distance $d_{1i}$ between the nucleus $S_1$ and the destination station $S_i$.}
    \label{fig:monocentric}
\end{figure}

In reality, the data does not lie on the line given by $\hat{\mu}_i = d{1i}$, as shown in Figure \ref{fig:monocentric}. In the Figure, the network distance from the nucleus to the destination station is represented on the $x$-axis and the average length of journeys arriving at a station is represented on the $y$-axis. Each bubble in the plot represents one station. The solid line is the regression line obtained via ordinary least squares. The results of the linear regression are shown in Table \ref{tab:monocentric}. The red dotted line represents equation \eqref{hypothesis}, i.e. the line where points would lie if the \textit{monocentric hypothesis} was satisfied. In fact, there is a tendency for the average length of the journeys terminating at station $S_i$ to be less than $d_{1i}$ as $d_{1i}$ gets larger. This suggests that journeys which terminate at stations that are far from the nucleus, tend to take place more locally. The effect is particularly obvious in Seoul, showing that the observed patterns of mobility depart from the \textit{monocentric hypothesis} to a greater extent. 

\renewcommand{\arraystretch}{1.5}
 \begin{table}[ht]
 \centering 
 \captionsetup{width=\textwidth}
 \caption{Results of linear regressions considering $d_{1i}$ as the explanatory variable and $\hat{\mu}_i$ as the response variables. Piccadilly Circus is considered to be the nucleus in London and City Hall in Seoul.} 
 \begin{tabular}{c|cccc}
  \multicolumn{1}{c}{} & Intercept & Slope & $R$ & $p$-value\\
  \hline
  \textbf{London}  & $4.352 \pm 0.283$ & $0.580 \pm 0.027$ & $0.758$ & $<0.05$\\
  \hline
  \textbf{Seoul} & $8.203 \pm 0.157$ & $0.152 \pm 0.010$ & $0.585$ & $<0.05$\\
 \end{tabular}\label{tab:monocentric}
\vspace{5mm}
\end{table}
\renewcommand{\arraystretch}{1}

\begin{figure}[ht!]
\centering
\leavevmode
    \hbox{\includegraphics[width=0.58\textwidth]{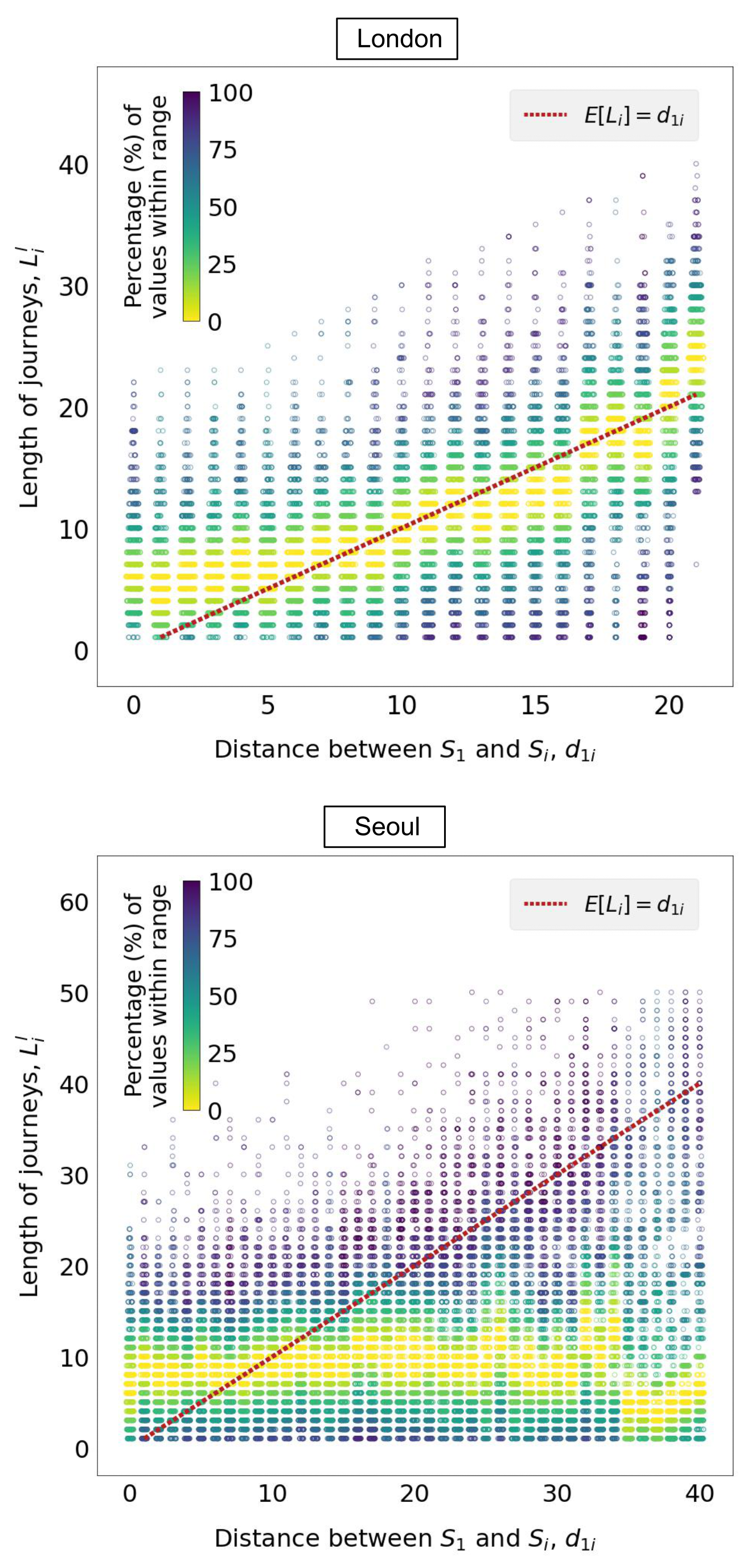}}
    \vspace*{3mm}
    \caption{Distribution of the length of journeys terminating at any destination station which is at a given distance from the nucleus. The solid line indicates the median of the distribution for each value of $d_{1i}$. The dashed line represents the line $L_i^l = d_{1i}$.}
    \label{fig:individual_journeys}
\end{figure}

In addition, we observe not only that $\hat{\mu}_i = d{1i}$ is not satisfied, but also that $L_i$ displays a high degree of variability for $i=1,...,N$. This effect is captured in Figure \ref{fig:individual_journeys}, where each data point corresponds to an individual journey, the $x$-coordinate represents the distance $d_{1i}$ between the nucleus and the destination station, and the $y$-coordinate represents the length of each individual journey $L_i^l$, with $l=1,...,M_i$ and $i=1,...,N$. Once again, the red dotted line represents equation \eqref{hypothesis}.

Figures \ref{fig:monocentric} and \ref{fig:individual_journeys} are thus a manifestation that real cities do not conform to the hypothesised monocentric scenario. Next, explore the deviations from the \textit{monocentric hypothesis} by leveraging the observed variability in our data.

\subsection{An approach based on mixture models}

In order to describe the deviations from the hypothesised monocentric behaviour, we introduce mixture models, which are probabilistic models with the ability to represent the possible presence of different statistical sub-populations within the overall population. In the context of this paper, mixture models can be used to infer the possible presence of centres other than the nucleus based only on the data for the number and length of journeys terminating at each station. The approach that we propose here consists in assuming that the true probability distribution for the number of journeys to station $S_i$ is given by a mixture distribution of the following form 

\begin{equation}\label{mixture_general}
    f_i(L_i = h|\mathbf{w}_i, \boldsymbol\theta_i) = \sum_{j=1}^Kw_i^jp_i^j(L_i = h|\boldsymbol\theta_i^j).
\end{equation}

In equation \eqref{mixture_general}, the probability that a journey terminating at station $S_i$ has length $L_i = h$, is now conditional on the parameters of the true distribution, $\mathbf{w}_i$ and $\boldsymbol\theta_i$. The probability density function of the true distribution is given by a weighted sum of $K$ probability density functions corresponding to each of the components of the mixture. The number of components in the mixture $K$ corresponds to the number of centres assumed by the model. If $K=1$, then the only centre accounted for in the model is the nucleus, but if $K>1$, the model assumes that there are subcentres other than the nucleus. The weights of these components are given by the column vector $\mathbf{w}_i$, with $K$ entries that satisfy $\sum_{j=1}^Kw^j_i = 1$. Each component has an associated probability density function given by $p^j_i(L_i=h|\boldsymbol\theta_i^j)$, with parameters $\boldsymbol\theta_i^j$ and so, $\boldsymbol\theta_i$ is a matrix with $K$ rows and as many columns as the number of parameters that characterise the probability density function $p^j_i$.

For the purposes of our analysis, we set $p^j_i$ to be a Poisson distribution with parameter $\mu_i^j$, so that $p^j_i(L_i=h|\boldsymbol\theta_i^j)$ is now $p^j_i(L_i=h|\mu_i^j) = \frac{1}{h!}(\mu_i^j)^h\exp(-\mu_i^j)$, for $i=1,...,N$ and $j=1,...,K$. This choice of distribution is motivated by the fact that the length of journeys terminating at station $S_i$ represented by random variable $L_i$ is in the form of count data, but also by the mathematical simplicity of the Poisson distribution, which is characterised by only one parameter.

In order to find the maximum likelihood estimates of $\mathbf{w}_i$ and $\boldsymbol\theta_i$ given the observed data for station $S_i$, we apply the expectation-maximisation algorithm. The number of components $K$ for the Poisson mixtures is one of the hyperparameters of the algorithm and needs to be determined before the learning process. In section \ref{sec:results}, we discuss different choices for the number of components $K$.

\section{Experiments and results}\label{sec:results}

The case where $K=1$ is equivalent to a simple Poisson distribution. The maximum likelihood estimator for the Poisson parameter $\mu_i$ corresponding to station $S_i$ is the average of the $M_i$ observations for $L_i$. Therefore, a visualisation of the relation between the estimated Poisson parameter corresponding to a station and the distance between the nucleus and the station is provided in Figure \ref{fig:monocentric}.

\subsection{Poisson mixture model with two components}

To account for the presence of centres other than the nucleus, we introduce Poisson mixture models with $K=2$. The parameters of a Poisson mixture with two components are the weights $\mathbf{w}_i = (w^1_i, w^2_i)$ and the distribution parameters for each component $\boldsymbol\theta_i = (\mu^1_i, \mu^2_i)$, which are also the mean of each component. We can obtain the maximum likelihood estimators for these parameters by applying the expectation-maximisation algorithm to data corresponding to $L_i$ and we denote them by $\hat{w}^1_i$, $\hat{w}^2_i$, $\hat{\mu}^1_i$ and $\hat{\mu}^2_i$. We refer to the component with the lowest estimated mean as the proximal component. We denote its associated weight by $\hat{w}^p_i$ and its mean by $ \hat{\mu}^p_i $, so that $ \hat{\mu}^p_i = min( \hat{\mu}^1_i,  \hat{\mu}^2_i)$. Similarly, we call the distal component the component whose estimated mean is higher and denote its associated weight by $\hat{w}^d_i$ and its mean by $ \hat{\mu}^d_i $, so that  $ \hat{\mu}^d_i = max( \hat{\mu}^1_i, \hat{\mu}^2_i)$. The 2-component Poisson mixture model enables us to classify each of the $M_i$ individual observations of $L_i$ as belonging to the proximal or distal components with probability given by the respective estimated weights. Therefore, journeys belonging to the proximal component are those that take place at a local scale and the journeys belonging to the distal component take place at a global, city-wide scale. 

\begin{figure} [ht]
    \centering
    \leavevmode\hbox{\includegraphics[width=0.68\textwidth]{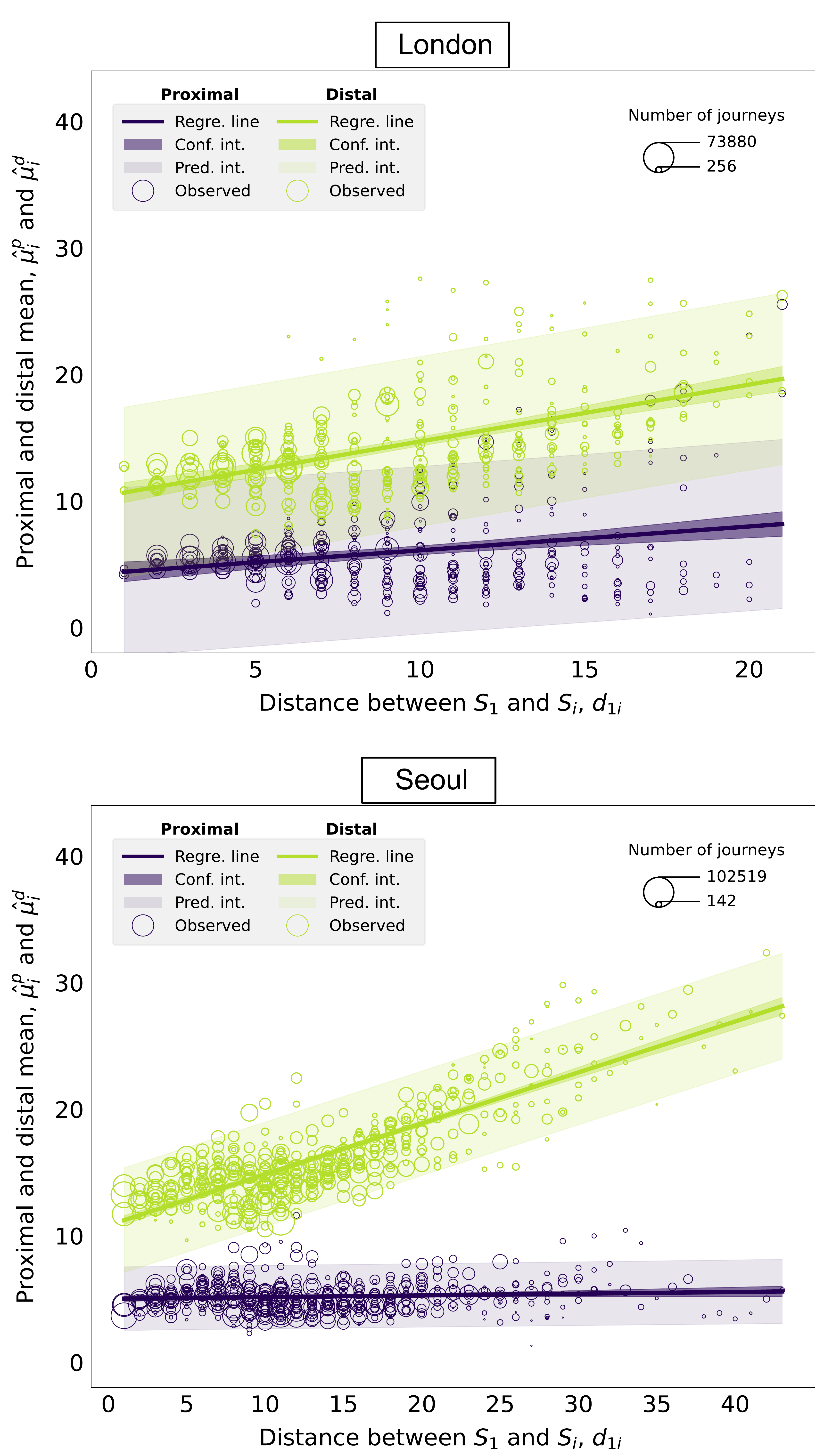}}
    \vspace*{3mm}
    \caption{Relationship between the proximal and distal mean of the distribution of journeys terminating at each station, $\hat{\mu}^p_i$ and $\hat{\mu}^d_i$, and the distance $d_{1i}$ between the nucleus $S_1$ and the destination station $S_i$.}
    \label{fig:2mixture}
\end{figure}

Figure \ref{fig:2mixture} shows the relationship between the proximal and distal means corresponding to a given station $S_i$ and the network distance $d_{1i}$ between $S_i$ and the nucleus $S_1$. Both the proximal and distal means, $\hat{\mu}^p_i$ and $ \hat{\mu}^d_i$, are represented in the $y$-coordinate. The size of the data points is proportional to the number of journeys terminating at each station. The results of the regression are displayed in Table \ref{tab:2_mixture}. In the Supplementary Information, we discuss the relationship between $d_{1i}$ and the weights corresponding to station $S_i$, $\hat{w}^d_i$, $\hat{w}^p_i$.

As $d_{1i}$ becomes larger, there is no significant increase in the proximal mean $\hat{\mu}^p_i$, since it remains around 5 and never above 10. The effect is strikingly obvious in the case of Seoul. These observations are likely to be the consequence of the existence of other socioeconomic centres, closer to the destination station $S_i$, where passengers prefer to travel to carry out some socioeconomic activities at a more local level. In contrast, the distal component displays a significant linear growth with $d_{1i}$. The distal component captures long-distance, city-wide journeys from stations that are possibly close to the nucleus, to stations that are in the peripheral regions of the city.

 \begin{table}[ht]
 \centering 
 \captionsetup{width=\textwidth}
 \caption{Results of linear regressions considering $d_{1i}$ as the explanatory variable and $\hat{\mu}_i^p$, $\hat{\mu}_i^d$ from the 2-component Poisson mixture model as the response variables. Piccadilly Circus is considered to be the nucleus in London and City Hall in Seoul.}
 \vspace{8mm}
 \begin{tabular}{cc|cccc}
  \multicolumn{2}{c}{} & Intercept & Slope & $R$ & $p$-value\\
  \hline
  \multirow{2}{*}{\textbf{London}} & Proximal & $4.295 \pm 0.437$ & $0.180 \pm 0.041$ & $0.228$ & $<0.05$ \\
   & Distal & $10.348 \pm 0.435$ & $0.427 \pm 0.041$ & $0.488$ & $<0.05$ \\
  \hline
  \multirow{2}{*}{\textbf{Seoul}}& Proximal & $5.018 \pm 0.116$ & $0.013 \pm 0.007$ & $0.086$ & $0.06$ \\
  & Distal & $10.826 \pm 0.191$ & $0.403 \pm 0.012$ & $0.844$ & $<0.05$ \\
 \end{tabular}\label{tab:2_mixture}
\vspace{5mm}
\end{table}

\subsection{Poisson mixture model with three components}

\begin{figure} [ht]
    \centering
    \leavevmode\hbox{\includegraphics[width=0.68\textwidth]{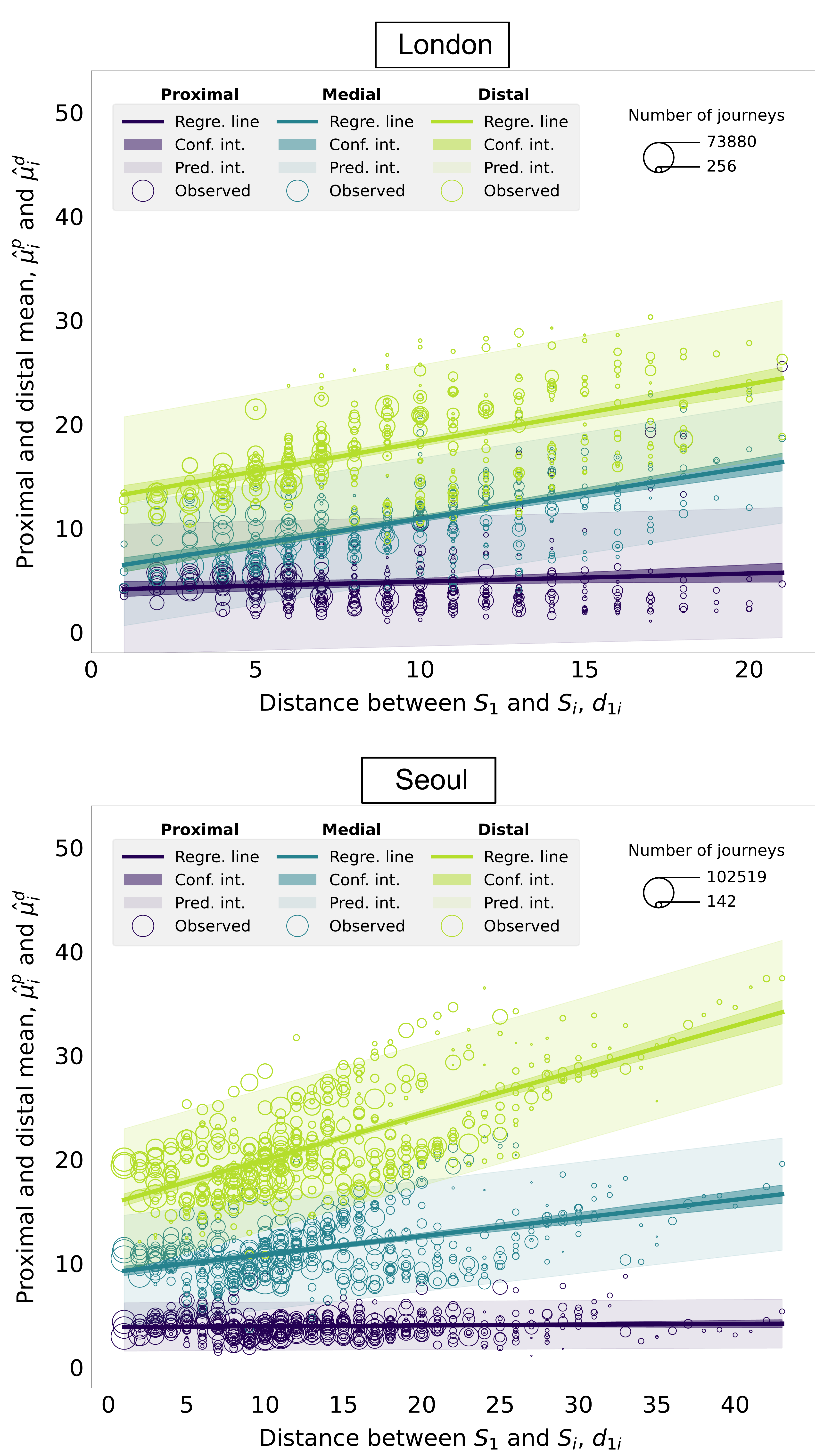}}
    \vspace*{3mm}
    \caption{Relationship between the proximal, medial and distal mean of the distribution of journeys terminating at each station, $\hat{\mu}^p_i$, $\hat{\mu}^m_i$ and $\hat{\mu}^d_i$, and the distance $d_{1i}$ between the nucleus $S_1$ and the destination station $S_i$.}\label{fig:3mixture}
\end{figure}

Similarly, with $K=3$, the parameters to be estimated are $\mathbf{w}_i = (w^1_i, w^2_i, w^3_i)$ and $\boldsymbol\theta_i = (\mu^1_i, \mu^2_i, \mu^3_i)$. The proximal and distal components are defined as for $K=2$. We call the third component remaining the medial component, and we denote its weight by $w^m_i$ and its distribution parameter by $\mu^m_i$. The results of the linear regression between the mean of each component and $d_{1i}$ are gathered in Table \ref{tab:3_mixture}. Figure \ref{fig:3mixture} represents the relationship between the proximal, medial and distal means corresponding to each station $S_i$, i.e. $\mu^p_i$, $\mu^m_i$ and $\mu^d_i$, and the distance between $S_i$ and the nucleus. In the Supplementary Information, we discuss the relationship between $d_{1i}$ and the weights corresponding to the three-component Poisson mixture model for station $S_i$, i.e. $\hat{w}^d_i$, $\hat{w}^d_i$, $\hat{w}^p_i$.

The behaviour of the proximal and distal components is analogous to the $K=2$ case. However, adding a third component allows to capture the variability in the data with even more detail. Theoretically, Poisson mixture models have no limitation for the maximum number of components to be added in their formulation, however, increasing $K$ indefinitely is not always sensible since it could lead to overfitting and hinder the interpretability of outcomes. For this reason, here we recommend keeping $K$ to 2 or 3 as a good trade-off between capturing the detail in the data variability whilst keeping the components meaningful without overcomplicating the model. 

\vspace{4mm}
 \begin{table}[ht]
 \centering 
 \captionsetup{width=\textwidth}
 \caption{Results of linear regressions considering $d_{1i}$ as the explanatory variable and $\hat{\mu}_i^p$, $\hat{\mu}_i^m$, $\hat{\mu}_i^d$ from the 3-component Poisson mixture model as the response variables. Piccadilly Circus is considered to be the nucleus in London and City Hall in Seoul.}\label{tab:3_mixture}
 \vspace{8mm}
 \begin{tabular}{cc|cccc}
  \multicolumn{2}{c}{} & Intercept & Slope & $R$ & $p$-value\\
  \hline
  \multirow{2}{*}{\textbf{London}} & Proximal & $4.070 \pm 0.407$ & $0.079 \pm 0.038$ & $0.109$ & $<0.05$ \\
   & Medial & $5.982 \pm 0.381$ & $0.494 \pm 0.036$ & $0.593$ & $<0.05$ \\
   & Distal & $12.683 \pm 0.487$ & $0.558 \pm 0.046$ & $0.546$ & $<0.05$ \\
  \hline
  \multirow{2}{*}{\textbf{Seoul}} & Proximal & $3.876 \pm 0.107$ & $0.008 \pm 0.007$ & $0.052$ & $0.252$ \\
   & Medial & $9.108 \pm 0.246$ & $0.175 \pm 0.015$ & $0.468$ & $<0.05$ \\
   & Distal & $15.678 \pm 0.315$ & $0.430 \pm 0.091$ & $0.713$ & $<0.05$ \\
 \end{tabular}
\vspace{5mm}
\end{table}

\section{Discussion and conclusions}

Our work constitutes a novel approach to the study of urban structure that makes use of a probabilistic modelling framework based on Poisson mixture models and new forms of data. Simply by analysing data related to the length of the journeys taken on the public transport system corresponding to London, United Kingdom, and Seoul, South Korea, the proposed probabilistic modelling framework enables us to disaggregate the journeys into several statistical subpopulations according to their destination station and their length, measured on network distances between stations. The methodology relies on the \textit{monocentric hypothesis}, a null hypothesis stating that in a perfectly monocentric city, there is only one centre where all the socioeconomic activity is concentrated. This centre is set to be at the station with the largest number persons travelling to it associated with the number of tap-outs, which we call the nucleus. Consequently, in the \textit{monocentric hypothesis}, all the journeys terminating at any station other than the nucleus should have the nucleus as their origin station. Characterising deviations from the \textit{monocentric hypothesis} allows us to infer the degree of polycentricity of a city by detecting the potential presence of centres other than the nucleus.

The analysis of data from London and Seoul leads us to the following key findings. Firstly, the distribution of the length of journeys terminating at any destination station displays a higher degree of variability than predicted by the \textit{monocentric hypothesis} in both London and Seoul. Secondly, by modelling the length of journeys terminating at each station in the public transport system with a single-component Poisson distribution, we observe that, especially in the case of Seoul, the most frequent journey length terminating at a given station is shorter than the distance between the nucleus and the station, perhaps due to the presence of closer, more local urban centres. Thirdly, by introducing the 2-component Poisson mixture model, we are able to classify each of the individual journeys to a station into what we call the proximal and distal components. The proximal component corresponds to journeys that take place at a local scale and the distal component involves journeys that take place at a global, city-wide scale. We see that regardless of the distance between the destination station and the nucleus, the most frequently observed journey length associated with the proximal component is around 5, as measured by network distance or number of stops away. These observations are particularly clear in the case of Seoul and are likely to be the consequence of the existence of other socioeconomic centres, closer to the destination station, where passengers may prefer to travel to carry out some socioeconomic activities at a more local level. Conversely, the most frequently observed journey length associated with the distal component is larger than the distance from the nucleus to the destination station for destination stations that are close to the nucleus, showing that passengers who terminate their journey at one of these stations may be travelling not only from the nucleus, but also from other origin stations that are further away. As the distance from the nucleus to the destination station increases, the most frequently observed journey length associated with the distal component increases fast, indicating that the distal component captures long-distance, city-wide journeys from origin stations that are possibly close to the nucleus, to stations that are in the peripheral regions of the city. Finally, increasing the number of components in the Poisson mixture model can help unpick the detail in the variability of the data; however, it can also make the model too complex and result in overfitting. After testing for other choices of nucleus (e.g. station with highest between-ness centrality in the transport network), we find that the observed patterns described above still hold. 

Understanding urban structure from data related to the transportation system of the cities has significant implications for urban policy. The modelling framework outlined in this paper provides a quantitative way of characterising the degree of monocentricity of a city and we suggest that this approach may be useful in the context of making decisions for spatial strategic planning. In particular, the recent quest for centralizing activities in more compact cities where the emphasis has been upon inner and city center living, could be much informed by this approach where the difficulty of moving towards more compact urban structures might be measured by the different parameters associated with the Poisson mixture distributions. In this sense, these distributions are not only able to reveal how the centricity of cities might change under different travel regimes but also how travel behavior itself might be altered. 

Additionally, our approach is also of relevance to those interested in the most theoretical, and even historical, aspects of urban areas and it prompts questions for future research. As our findings indicate, London's case study aligns more with the scenario depicted by the \textit{monocentric hypothesis} than Seoul's. The construction of London's transport network started at the end of the 19th century while the construction of Seoul's started in 1971, approximately a century later, and in the course of all those years, several studies have reported a tendency for cities to become more and more polycentric. Assuming that the layout of the transport network and the passengers' travelling behaviours are yet another manifestation of urban structure, then the fact that our findings suggest that London is more monocentric than Seoul, should not come as a surprise. But, does this assumption hold in general? Has London's early construction of a public transport network conditioned its urban structure and slowed its transition towards a more polycentric arrangement like Seoul's? There is considerable scope for extending this type of analysis to the evolution of past cities, developing where possible ways in which mixture models like these can reveal how spatial behaviors can and have altered over decades. Public transport data will always be an issue for such historical analysis but these methods can easily be extended to other trip distributions such as the journey to work on different modes and very different time intervals where there is data available in the UK for example, for the last 6 decades.

This study is limited by the fact that the data set is relatively small, since it only contains data for two cities and for five weekdays in the case of London, and four in the case of Seoul. Similarly, the data is limited to some modes of transport in the whole public transport system of both cities. For example, it excludes bus journeys, although tap-out data for bus journeys is not always recorded. Finally, our work only explores one temporal scale, but a deeper understanding of urban structure could be obtained by studying data at different times of the day as well as time periods of different lengths, and by analysing the temporal evolution of the data.

\bibliography{references.bib}

\section*{Acknowledgements}

We are very grateful to Prof HaeRan Shin from Seoul National University for kindly helping us get access to T-money data. CZ, MB and RS are Turing Fellows at the Alan Turing Institute. The data analysis was done within UCL.

\section*{Competing interests}

The authors declare that they have no competing interests.

\section*{Author's contributions}

CCA: conceived the study, designed the methodology, carried out the data analysis, contributed towards the interpretation of results, drafted the manuscript; CZ: conceived the study, provided the London data, contributed towards the interpretation of results; MB: contributed towards the interpretation of results, participated in drafting and revising the manuscript; RS: conceived the study, designed the methodology, contributed towards the interpretation of results; SMK: conceived the study, provided the Seoul data, contributed towards the interpretation of results. All authors gave their final approval for publication.

\section*{Data and code availability}

The data and the code used in the methodology of this paper can be found on GitHub through this \href{https://github.com/CrmnCA/inferring-urban-polycentricity-from-variability-in-human-mobility}{link}.

\section*{Funding}

This project has received funding from the European Research Council (ERC) under the European Union’s Horizon 2020 research and innovation programme (grant agreement No. 949670), and from ESRC under JPI Urban Europe/NSFC (grant No. ES/T000287/1).

\end{document}